\documentclass[a4paper,11pt]{article} 
\usepackage{cite}
\usepackage{setspace}
\usepackage{ae,aecompl,amsbsy,amssymb,eurosym}
\usepackage[english]{babel}
\usepackage{graphicx}
\usepackage{epsfig}
\usepackage{epstopdf}
\usepackage{amsmath}
\usepackage{amssymb}
\usepackage{subfigure}
\usepackage{ifthen}
\usepackage{alltt}
\usepackage{fancyhdr}
\usepackage{verbatim}
\usepackage{moreverb}
\usepackage{url}

\newcommand{\vt}[1]{\ensuremath{\mathbf{#1}}} 
\newcommand{\lt}[1]{\ensuremath{\mathrm{#1}}} 

\newcommand{\unitx}{\ensuremath{\vt{u}_x}}
\newcommand{\unity}{\ensuremath{\vt{u}_y}}
\newcommand{\unitz}{\ensuremath{\vt{u}_z}}

\newcommand{\unitr}{\ensuremath{\vt{u}_r}}
\newcommand{\unitt}{\ensuremath{\vt{u}_\theta}}
\newcommand{\unitp}{\ensuremath{\vt{u}_\phi}}

\newcommand{\unit}{\ensuremath{\vt{u}}}

\newcommand{\adyad}{\ensuremath{\overline{\overline{\alpha}}}}

\newcommand{\unitdyad}{\ensuremath{\overline{\overline{I}}}}

\fancypagestyle{plain}{
  \fancyhead{} 
}

\begin{document}

\author{Antti~O.~Karilainen and Sergei~A.~Tretyakov\\
\\
Department of Radio Science and Engineering\\
/ SMARAD Center of Excellence,\\
Aalto University, P.~O.~Box 13000, FI-00076 Aalto, Finland.\\
Email: antti.karilainen@aalto.fi.}

\title{Isotropic Chiral Objects With Zero Backscattering}

\maketitle

\begin{abstract}
In this paper we study electrically small chiral objects with isotropic response and zero backscattering. A bi-isotropic sphere is used as a simple example and its zero-backscattering conditions are studied. A theoretical model of an object composed of three orthogonal chiral particles made of conducting wire is presented as an analog of the zero-backscattering bi-isotropic sphere. A potential application of the object as a receiving antenna or a sensor with the ability to receive power from an arbitrary direction without backscattering is discussed.
\end{abstract}



\section{Introduction}

Objects invisible when observed from any direction as well as the concept of cloaking have aroused interest in the literature, but more applications are available if we allow objects to scatter power. If an object absorbs some power, which is the case of receiving antennas and sensors, it is dictated by the optical theorem that some power is scattered in the forward direction \cite{Bowman1969}. Therefore, we must settle for forward scattering, but are free to reduce the scattering in the opposite direction.

Finite objects with zero backscattering are known from the literature, a classical example being an object with equal relative material parameters, $\mu_\lt{r} = \epsilon_\lt{r}$, and $\pi/2$ rotational symmetry when observed from the incident direction \cite{Wagner1963,Weston1963}. Recently, it has been shown that in addition to the $\pi/2$ rotational symmetry, self-dual nature of the object is sufficient for zero backscattering \cite{Lindell2009}. For example, a DB sphere was studied in \cite{Sihvola2009}, and a D'B' sphere and cube in \cite{Lindell2009}. Further, it was shown in \cite{Lindell2009} that the physical geometry of the object does not need to have the $\pi/2$ rotational symmetry, as long as the geometrical asymmetry is balanced by anisotropy of material response.

An electrically small object formed by two crossed chiral particles was studied in \cite{Karilainen2011,Karilainen2011c,Karilainen2011e}, and it was shown that such object does not backscatter any power if the incidence direction is normal to the plane of the particles, even if the object receives power. The combination of polarizabilities of the chiral particles is self dual with the needed symmetry, which explains the phenomena. If the combination of chiral particles is an electrically small object with induced dipole moments dominating over higher-order moments, the forward-scattering pattern is that of a Huygens' pattern with one null in the back direction. The set of two chiral particles is limited in the sense that zero backscattering happens only in two directions, normal to the plane of the particles. In this paper we consider isotropic objects with zero backscattering that have the same response in every direction.

We complement the set of two orthogonal chiral particles to an isotropic chiral object by adding a third chiral particle perpendicular to the original particles. A simple object analogous to the proposed isotropic chiral object is a lossless bi-isotropic chiral sphere. Scattering from bi-isotropic spheres with chiral (optically active) magneto-electric coupling have been thoroughly studied in \cite{Bohren1974,Lakhtakia1989a,Lindell1990}.

In the following sections, zero-backscattering conditions for the bi-isotropic sphere are studied with emphasis on the forward scattered pattern and polarization. Next, the induced dipole moments are written and self-duality is shown for the isotropic chiral object. Backscattering and forward scattering from the object are solved and compared to the chiral sphere. Applications for the isotropic chiral object are discussed in the final section.


\section{Bi-Isotropic Sphere}
\label{sec:isotropic_sphere}

Scattered fields from an electrically small bi-anisotropic scatterer illuminated by an incident plane wave can be solved using its polarizabilities defined as
{\setlength\arraycolsep{2pt}
\begin{eqnarray}
	\vt{p} & = & \adyad_\lt{ee} \cdot \vt{E}_\lt{inc} + \adyad_\lt{em} \cdot \vt{H}_\lt{inc},
	\label{eq:psimple} \\
	\vt{m} & = & \adyad_\lt{me} \cdot \vt{E}_\lt{inc} + \adyad_\lt{mm} \cdot \vt{H}_\lt{inc},
	\label{eq:msimple}
\end{eqnarray}
where $\vt{p}$ and $\vt{m}$ are the induced electric and magnetic dipole moments, respectively, $\vt{E}_\lt{inc}$, $\vt{H}_\lt{inc}$ are the incident fields, and $\adyad$ are the polarizabilities \cite{Serdyukov2001}. For zero backscattering the polarizability dyadics in (\ref{eq:psimple}) and (\ref{eq:msimple}) need to be self dual \cite{Lindell2009}. One can carry out an analysis, similarly as for self-dual media in \cite{Lindell2009}, by demanding that the duality transformed (\ref{eq:psimple}) and (\ref{eq:msimple}) equal the originals, and arrive at the conditions
{\setlength\arraycolsep{2pt}
\begin{eqnarray}
	\adyad_\lt{mm} & = & \eta_0^2 \adyad_\lt{ee}, \label{eq:mm_ee} \\
	\adyad_\lt{em} & = & -\adyad_\lt{me}, \label{eq:em_me}
\end{eqnarray}
where $\eta_0 = \sqrt{\mu_0 /\epsilon_0}$ is the wave impedance of the surrounding space. The first equation (\ref{eq:mm_ee}) tells that the electric and magnetic responses to electric and magnetic fields, respectively, are equal. The second sufficient condition for zero backscattering is the rotational symmetry of $\pi/2$~\cite{Lindell2009}. Because we are interested in zero backscattering from any direction, let us therefore focus on isotropic objects, that are a natural choice if no certain direction is special. Further, we use the bi-isotropic sphere as an example in the following, because it obviously has the needed symmetry in all directions.

For bi-isotropic scatterers the polarizabilities in (\ref{eq:psimple}) and (\ref{eq:msimple}) are of course not anymore dyadics but scalar parameters. The polarizabilities of a bi-isotropic sphere with radius $a$ and volume $V = \frac{4}{3}\pi a^3$ with uniform incident electric and magnetic fields are~\cite{Sihvola1992}
{\setlength\arraycolsep{2pt}
\begin{eqnarray}
	\alpha_\lt{ee} & = & 3V \epsilon_0 \frac{
	  (\mu_\lt{r} + 2)(\epsilon_\lt{r} - 1) - (\chi^2 + \kappa^2)
	  }{ 
	  (\mu_\lt{r} + 2)(\epsilon_\lt{r} + 2) - (\chi^2 + \kappa^2)
	  }, \label{eq:aee_sphere} \\
	 \alpha_\lt{em} & = & \frac{3V}{\sqrt{\mu_0 \epsilon_0}} \frac{
	  3 (\chi - j\kappa)
	  }{ 
	  (\mu_\lt{r} + 2)(\epsilon_\lt{r} + 2) - (\chi^2 + \kappa^2)
	  }, \label{eq:aem_sphere} \\
	 \alpha_\lt{me} & = & \frac{3V}{\sqrt{\mu_0 \epsilon_0}} \frac{
	  3(\chi + j\kappa)
	  }{ 
	  (\mu_\lt{r} + 2)(\epsilon_\lt{r} + 2) - (\chi^2 + \kappa^2)
	  }, \label{eq:ame_sphere} \\
	 \alpha_\lt{mm} & = & 3V \mu_0 \frac{
	  (\mu_\lt{r} - 1)(\epsilon_\lt{r} + 2) - (\chi^2 + \kappa^2)
	  }{ 
	  (\mu_\lt{r} + 2)(\epsilon_\lt{r} + 2) - (\chi^2 + \kappa^2)
	  }, \label{eq:amm_sphere}
\end{eqnarray}
where $\mu_\lt{r}$ is the relative permeability, $\epsilon_\lt{r}$ the relative permittivity, $\kappa$ the chirality parameter, and $\chi$ the non-reciprocity parameter.

Condition (\ref{eq:mm_ee}) is satisfied in (\ref{eq:aee_sphere}) and (\ref{eq:amm_sphere}) for any values of $\chi$ and $\kappa$ if $\mu_\lt{r} = \epsilon_\lt{r}$. With $\chi = 0$, $\kappa = 0$ we arrive at the classical case of a zero-backscattering object with material parameters $\mu_\lt{r} = \epsilon_\lt{r}$. By relating the second condition (\ref{eq:em_me}) to (\ref{eq:aem_sphere}) and (\ref{eq:ame_sphere}) we see that $\chi$ must vanish. The conclusion is that a bi-isotropic zero-backscattering sphere is a reciprocal isotropic chiral sphere, with an arbitrary value of the chirality parameter $\kappa$. This zero-backscattering condition for chiral bodies of revolution appears also in~\cite{Uslenghi1990,Uslenghi1996}, expressed in terms of a different set of material parameters.

As chirality is not a necessary condition for zero backscattering, it is of interest to consider what kind of an effect it brings into the scattered fields in the forward direction. The induced dipole moments in the quasistatic limit (electrical size $ka \ll 1$, $k = \omega\sqrt{\mu_0 \epsilon_0}$) for a $z$-directed and $x$-polarized plane wave are~\cite{Lindell1990}
{\setlength\arraycolsep{0pt}
\begin{eqnarray}
	\vt{p} = 4\pi \epsilon_0 & & E_0 a^3 \bigg \{ \frac{
	  (\mu_\lt{r} + 2)(\epsilon_\lt{r} - 1) - (\chi^2 + \kappa^2)
	  }{ 
	  (\mu_\lt{r} + 2)(\epsilon_\lt{r} + 2) - (\chi^2 + \kappa^2)
	  } \unitx \nonumber \\
	  & & + \frac{
	  3(\chi - j\kappa)
	  }{ 
	  (\mu_\lt{r} + 2)(\epsilon_\lt{r} + 2) - (\chi^2 + \kappa^2)
	  } \unity
	  \bigg \}, \label{eq:p_sphere} \\
	 \vt{m} = 4\pi \epsilon_0 & & E_0 \eta_0 a^3 \bigg \{ \frac{
	  3(\chi + j\kappa)
	  }{
	  (\mu_\lt{r} + 2)(\epsilon_\lt{r} + 2) - (\chi^2 + \kappa^2)
	  } \unitx \nonumber \\
	  & & + \frac{
	  (\mu_\lt{r} - 1)(\epsilon_\lt{r} + 2) - (\chi^2 + \kappa^2)
	  }{ 
	  (\mu_\lt{r} + 2)(\epsilon_\lt{r} + 2) - (\chi^2 + \kappa^2)
	  } \unity
	  \bigg \}, \label{eq:m_sphere}
\end{eqnarray}
where $E_0$ is the magnitude of the incident electric field. The effect of chirality is now easy to identify from (\ref{eq:p_sphere}) and (\ref{eq:m_sphere}): the relation $\mu_\lt{r} = \epsilon_\lt{r}$ leads to co-polarized forward scattering with $p_x$ and $m_y$, but additional scattered components with $\kappa \neq 0$ contributes to a cross-polarized field via $p_y$ and $m_x$ \cite{Lindell1990}. Further, as the cross-polarized dipole moments have a $\pi/2$ phase shift with respect to the co-polarized component, chirality creates an elliptic component in the scattered field, which can be controlled without affecting the zero-backscattering condition.


\section{Isotropic Chiral Object From Chiral Particles}
\label{sec:isotropic_chiral}


\subsection{Induced Dipole Moments}
\label{sec:particles}

The chiral particles studied here are so-called idealized helices \cite{Jaggard1979}. One such particle is seen in Fig.~\ref{fig:canonical_particle} with corresponding parallel electric and magnetic dipole moments.
\begin{figure}[!t]
	\centering
	\includegraphics[width=30mm]{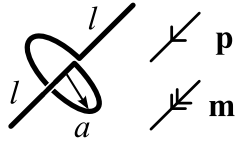}
	\caption{Right-handed chiral particle with parallel induced electric and magnetic dipoles.}
	\label{fig:canonical_particle}
\end{figure}
The polarizabilities of a single chiral particle are presented in \cite{Tretyakov1996}, but here we study, as in \cite{Karilainen2011e}, the scattering from chiral particles, neglecting the electric polarizability of the loop. Using this approximation, only four polarizability components are left for each particle. For three identical $x$, $y$ and $z$ directed chiral particles with the same handedness the polarizabilities are
{\setlength\arraycolsep{2pt}
\begin{eqnarray}
	\adyad_\lt{ee} & = & a_\lt{ee} \unitdyad, \nonumber \\
	\adyad_\lt{em} & = & a_\lt{em} \unitdyad, \nonumber \\
	\adyad_\lt{me} & = & a_\lt{me} \unitdyad, \nonumber \\
	\adyad_\lt{mm} & = & a_\lt{mm} \unitdyad,	\label{eq:pols}
\end{eqnarray}
where $\unitdyad = \unitx\unitx + \unity\unity + \unitz\unitz$ is the unit dyadic. The polarizabilities for each particle, with the balance condition for the scattered power $Sk = l$ \cite{Alitalo2011,Karilainen2011e}, where $l$ is the length of the wire dipole arm and $S = \pi a^2$ is the area of the loop (see Fig.~\ref{fig:canonical_particle}), are
\begin{eqnarray}
	\alpha_\lt{ee} & = & \frac{l^2}{j\omega}\frac{1}{Z_\lt{a}}, \label{eq:a_ee} \\	
	\alpha_\lt{em} & = & \pm \eta_0 \frac{l^2}{\omega}\frac{1}{Z_\lt{a}} = \pm j\eta_0 a_\lt{ee} \label{eq:a_em}, \\
	\alpha_\lt{me} & = & -a_\lt{em} = \mp j\eta_0 a_\lt{ee}, \label{eq:a_me} \\
	\alpha_\lt{mm} & = & \eta_0^2 \frac{l^2}{j\omega}\frac{1}{Z_\lt{a}} = \eta_0^2 a_\lt{ee}, \label{eq:a_mm}
\end{eqnarray}
where $Z_\lt{a}$ is total impedance of the particles. In (\ref{eq:a_me}) the equality is due to reciprocity and in (\ref{eq:a_em}) the ``$\pm$'' sign defines the handedness of the particles: the top sign for left handed (LH) and the bottom sign for right handed (RH) here and in the following.

It should be noted that the balanced scattering polarizabilities (\ref{eq:a_ee})-(\ref{eq:a_mm}) can be in principle obtained with other chiral geometries than the one used in \cite{Tretyakov1996}, such as helices. In the following, we study these general balanced polarizabilities, thus not restricting the results to a certain chiral geometry.

Let us define an arbitrary incident plane wave electric field as
\begin{equation}
	\vt{E}_\lt{inc} = E_\lt{inc} \unit_0 e^{-j \vt{k} \cdot \vt{r}}
	\label{eq:Einc}
\end{equation}
where $E_\lt{inc}$ is constant, $\unit_0$ defines the polarization, the propagation direction is $\vt{u}$, and $\vt{k} = \vt{u} k$. The corresponding magnetic field is%
\begin{equation}
	\vt{H}_\lt{inc} = \frac{\vt{k}}{\omega \mu_0} \times \vt{E}_\lt{inc} = \frac{E_\lt{inc}}{\eta_0} \unit \times \unit_0 e^{-j \vt{k} \cdot \vt{r}}
	\label{eq:Hinc}
\end{equation}
The total induced dipole moments at the origin are written for this incident electric field using (\ref{eq:a_ee})-(\ref{eq:a_mm}) as follows:
{\setlength\arraycolsep{2pt}
\begin{eqnarray}
	\vt{p} & = & a_\lt{ee} E_\lt{inc} \unitdyad \cdot \unit_0  \pm j a_\lt{ee} E_\lt{inc} \unitdyad \cdot \unit \times \unit_0 ,
	\label{eq:p_isotropic} \\
	\vt{m} & = & \mp j \eta_0 a_\lt{ee} E_\lt{inc} \unitdyad \cdot \unit_0 + \eta_0 a_\lt{ee} E_\lt{inc} \unitdyad \cdot \unit \times \unit_0 .
	\label{eq:m_isotropic}
\end{eqnarray}
By examining (\ref{eq:p_isotropic}) and (\ref{eq:m_isotropic}), we can write
\begin{equation}
	\vt{m} = \mp j \eta_0 \vt{p}.
	\label{eq:dual_sources}
\end{equation}
We can see that the object is self dual when we compare (\ref{eq:a_mm}) to (\ref{eq:mm_ee}) and (\ref{eq:a_em}) and (\ref{eq:a_me}) to (\ref{eq:em_me}). In fact, (\ref{eq:dual_sources}) is the known relation between self-dual electric and magnetic sources~\cite{Lindell1996}.


\subsection{Scattering From the Object}
\label{sec:scattering}

As the object corresponding to three chiral particles was shown to be self dual, $\pi/2$ rotational symmetry is still required. If one draws three chiral particles or helices along $x$, $y$ and $z$ directions, one does not see such rotational symmetry in all directions. However, the symmetry requirement holds for the polarizabilities and the induced dipole moments.

The scattered electric far field from electric and magnetic dipole moments at the origin can be written, following \cite{Tai1971,Lindell2009}, as
\begin{equation}
	\vt{E}_\lt{sca} = -\omega^2 \mu_0 \frac{e^{-jkr}}{4 \pi r} \vt{F}_\lt{sca},
	\label{eq:Esca}
\end{equation}
where the radiation vector is
\begin{equation}
	\vt{F}_\lt{sca} = \unitr \times (\unitr \times \vt{p}) + \unitr \times \frac{\vt{m}}{\eta_0}.
	\label{eq:Fsca}
\end{equation}
Let us normalize the amplitude $a_\lt{ee} E_\lt{inc} \equiv 1$. Then, by inserting (\ref{eq:p_isotropic}) and (\ref{eq:m_isotropic}) to (\ref{eq:Fsca}), we get the radiation pattern as a function of the incident direction $\unit$ and the polarization vector $\unit_0$ as
\begin{equation}
	\vt{F}_\lt{sca}(\unit) = -(\unit_0 \pm j \unit \times \unit_0) \mp j \unitr \times (\unit_0 \pm j \unit \times \unit_0),
	\label{eq:Fsca_u}
\end{equation}
where the rotation identity $\unitr \times (\unitr \times \unit_0) = -\unit_0$ was used as $\unitr$ and $\unit_0$ are perpendicular.

The scattered field in the backscattering direction can be solved by substituting $\unitr = -\unit$, leading to
\begin{eqnarray}
	\vt{F}_\lt{sca}^\lt{BS} & = & -(\unit_0 \pm j \unit \times \unit_0) \mp j (-\unit) \times (\unit_0 \pm j \unit \times \unit_0) \nonumber \\
	& = & -\unit_0 \mp j\unit\times\unit_0 \pm j\unit\times\unit_0 - \unit\times\unit\times\unit_0 \nonumber \\
	& = & 0,
	\label{eq:Fsca_BS}
\end{eqnarray}
i.e.\ zero backscattering independent of the incident direction and polarization of the incident wave. Forward scattering is obtained by substituting $\unit = \unitr$ in (\ref{eq:Fsca_u}), and the result is
\begin{eqnarray}
	\vt{F}_\lt{sca}^\lt{FS} & = & -(\unit_0 \pm j \unit \times \unit_0) \mp j (\unit) \times (\unit_0 \pm j \unit \times \unit_0) \nonumber \\
	& = & -\unit_0 \mp j\unit\times\unit_0 \mp j\unit\times\unit_0 + \unit\times\unit\times\unit_0 \nonumber \\
	& = & -2(\unit_0 \pm j \unit \times \unit_0).
	\label{eq:Fsca_FS}
\end{eqnarray}
Hence, the forward scattering pattern is related to the polarization vector of the plane wave and the handedness of the particles.

In order to calculate the scattering pattern, let us fix the direction of the incident wave to $\unit = \unitz$ and write the incident electric field as
\begin{equation}
	\vt{E}_\lt{inc} = E_\lt{inc} \unit_0 e^{-jkz} = E_\lt{inc} (\unitx + jAR\unity) e^{-jkz}.
	\label{eq:Einc_z}
\end{equation}
where $AR$ is an optional elliptical component ($-1 \leq AR \leq 1$) and the sign of $AR$ denotes the handedness of the plane wave (plus sign for LH and minus sign for RH). The scattering pattern (\ref{eq:Fsca_u}) is written for forward scattering first in Cartesian coordinates with the standard transformation:
%
\begin{eqnarray}
	\vt{F}_\lt{sca}^\lt{FS} & = & -(1 \pm AR) \big[ \unitx (\cos\theta + 1) \pm j \unity (\cos\theta + 1) \nonumber \\
	& & - \unitz (\sin\theta\cos\phi \pm j\sin\theta\sin\phi) \big ].
	\label{eq:Fsca_step1}
\end{eqnarray}
Next, we transform $\unitx$, $\unity$ and $\unitz$ to spherical coordinates, neglecting the $\unitr$ component as we operate in the far zone, and write the complex pattern as
\begin{equation}
	\vt{F}_\lt{sca}^\lt{FS} = -(1 \pm AR)(\unitt\Theta_\lt{FS} + \unitp\Phi_\lt{FS}),
	\label{eq:Fsca_step2}
\end{equation}
where
\begin{eqnarray}
	\Theta_\lt{FS} & = & (\cos\theta + 1)(\cos\phi \pm j\sin\phi), \\
	\Phi_\lt{FS} & = & (\cos\theta + 1)(-\sin\phi \pm j\cos\phi).
	\label{eq:Fsca_step3}
\end{eqnarray}
Now the magnitude of the pattern is easy to calculate, and the result is
\begin{equation}
	|\vt{F}_\lt{sca}^\lt{FS}| = \sqrt{2}(1 \pm AR)(\cos\theta + 1),
	\label{eq:absFsca}
\end{equation}
which corresponds to the Huygens' pattern. The maximum field is scattered when the handedness of the incident wave and the particles are the same. On the other hand, the object is invisible if the handedness does not match. The axial ratio of the scattered field is easy to see from (\ref{eq:Fsca_FS}): its components are equal in magnitude and perpendicular to the propagation direction with a $\pi/2$ phase shift. Recalling the rotation direction of a plane wave, e.g., by comparing (\ref{eq:Fsca_FS}) to (\ref{eq:Einc_z}), the scattered field is always circularly polarized with the handedness determined by the handedness of the individual particles. The scattering properties from the balanced chiral particles correspond to those of the chiral sphere in Section~\ref{sec:isotropic_sphere} when $\kappa$ is chosen so that (\ref{eq:dual_sources}) holds for the chiral sphere.


\section{Discussion}
\label{sec:discussion}

Adding the third orthogonal chiral particle to a combination of two orthogonal particles with balanced chirality produces a self-dual isotropic chiral object. The self-dual properties of such particles were shown, but as it is hard to imagine an isotropic chiral object with $\pi/2$ rotational symmetry with a non-negligible electrical size, the physical realization is left for future study. Nonetheless, by examining the scattering from the set of self-dual dipole moments it was shown that the backscattering is zero for all incident directions and polarizations and the forward-scattered pattern is the Huygens' pattern. We notice that the (electrically small) isotropic chiral object composed only of conducting wires is an analog to the bi-isotropic chiral sphere in the quasistatic regime.

The chiral object combined with three receivers in each particle would be an ideal sensor for electromagnetic radiation. It was shown in \cite{Karilainen2011e} that two chiral particles can be conjugate loaded to receive all the available power according to the polarization matching coefficient while remaining in the zero-backscattering mode. Following the same reasoning, the isotropic object could have one electronically tunable receiver in each particle, thus allowing to synthesize the Huygens' reception pattern and to receive power from any direction with zero-backscattering. 

It is also possible to filter circular polarization using this scheme. If the incident wave is LH (RH) polarized, the object with RH (LH) chiral particles is invisible, but a matching handedness induces currents in the object, offering some control over the other circular polarization. Possible applications could be a reflection-less or an absorbing array with circular polarization filtering that could be realized by adding dissipation to the particles.

\section*{Acknowledgment}

A.~O.~Karilainen and S.~A.~Tretyakov acknowledge the support of the Academy of Finland and Nokia Corporation through the Center-of-Excellence program. The work of A.~O.~Karilainen has been supported by the Graduate School, Faculty of Electronics, Communications, and Automation; Elektro\-niikka\-insi\-n\"o\"orien s\"a\"ati\"o; and Nokia Foundation.




\end{document}